\documentclass[%
 preprint,longbibliography
 amsmath,amssymb,
 aps, physrev,
]{revtex4-2}

\usepackage{graphicx}
\usepackage{dcolumn}
\usepackage{bm}

\usepackage{epsf,amsmath,amssymb,url,dcolumn}
\usepackage[colorlinks=true,urlcolor=blue,linkcolor=blue,citecolor=blue]{hyperref}
\usepackage{mathtools}
\usepackage{color}
\usepackage[normalem]{ulem}
\usepackage{adjustbox}
\usepackage{array}
\usepackage{xcolor}
\usepackage{subcaption}
\usepackage{float}
\usepackage{tabularx}
\usepackage{makecell}
\usepackage{multirow}

\begin{document}

\preprint{APS/123-QED}

\title{\textbf{Model potential based estimation of positron bound states with polar molecules } 
}%

\author{Snigdha Sharma}
\author{Dhanoj Gupta}%
 \email{Contact author: dhanoj.gupta@vit.ac.in}
\affiliation{%
\\
Department of Physics, School of Advanced Sciences, Vellore Institute of Technology, Vellore, Tamil Nadu, India - 632014.
}%



\date{\today}

\begin{abstract}
Positron forming a bound state with molecules is an area of meticulous research and remains open to simpler techniques for estimating the binding energy. Correlation between the incoming positron and the molecular electrons plays a crucial role even for polar molecules, where binding is also possible at the static-potential level. In this work, we have used a model correlation potential, which accounts for the virtual positronium formation, in the single-center expansion scattering calculations for polar oxygen-containing molecules. The thus determined $S$-matrix is continued into the complex momentum plane by fitting it to the Padé approximant. The positron bound state is estimated using the calculated pole position of the $S$-matrix. The results, compared with the latest experimental and \textit{ab initio} findings, look quite promising.        

\end{abstract}

\maketitle


\textit{Introduction:} Interaction of positrons with matter and their annihilation finds various applications, including the vital positron emission tomography (PET) scans used to detect, diagnose, and monitor disease progression in oncology, neurology, as well as cardiology \cite{rong_radiochemistry_2023}. Enhanced positron annihilation rates, below the positronium formation threshold, in molecular gases were first observed by Paul and Saint-Pierre \cite{Paul_Pierre} and later evidenced by Gilbert \textit{et al.} \cite{Gilbert_2002} to be directly associated with vibrational resonances of the positron-molecule complex. Gilbert \textit{et al.}'s \cite{Gilbert_2002} success in measuring positron annihilation rates as a function of positron energy, spanning the vibrational energy levels, also provided direct evidence of its binding to molecules. 

Low-energy positrons can bind to molecules, with the excess energy transferred to the excitation of the molecular vibrational modes, leading to vibrational Feshbach resonances (VFRs). This bound resonant state of the positron-molecule complex greatly enhances the probability of annihilation. In experiments, the downshift of the observed VFR in the annihilation rate from the energy of the vibrational mode is used to determine positron-molecule binding energy ($\rm{\varepsilon_B}$) \cite{Gribakin_2010}. To date, $\rm{\varepsilon_B}$ for more than 100 molecules has been measured \cite{Danielson_2025}. The most successful theoretical estimation of $\rm{\varepsilon_B}$ involves the recent \textit{ab initio} many-body theory (MBT) approach, which factors in many-body correlations including the virtual positronium formation (temporary tunneling of molecular electron to the positron) \cite{hofierka2022many,Rawlins_2023,Cassidy_2024,Baidoo_2024}. Despite the promising results of the \textit{ab initio} method, the theoretical estimation of binding energies is still an open problem due to the computational challenges it involves \cite{Danielson_2025, fedus2025semi}. 

A relatively simpler technique includes the use of a model correlation potential and solving the Schrödinger equation to obtain the positron binding energy and its wavefunction, such as the DFT model implemented by Suzuki \textit{et al.} \cite{Suzuki_2020} and the free parameter model proposed by Swann and Gribakin  \cite{Swann_Gribakin_2018,Swann_Gribakin_2019,Swann_Gribakin_2020,Swann_2021}. They estimated the binding energy for both polar and non-polar molecules. Recent Schwinger multichannel (SMC) calculations \cite{Frighetto_2026, Frighetto_2024_Low} utilized the Swann and Gribakin model potential \cite{Swann_Gribakin_2018} and estimated the virtual and bound states by employing the near-threshold $s$-wave eigenphase approximation of Morrison \cite{Morrison}. This approach gave better results, with minimized computational cost, than the \textit{ab initio} SMC approach itself. In our previous work \cite{Snigdha_PRA}, we too incorporated the model correlation potential of Swann and Gribakin \cite{Swann_Gribakin_2018}---which takes care of the virtual positronium formation---into the single-center expansion (SCE) calculations, a model-potential-based method, and estimated the virtual and bound state energies by expanding the $s$-wave eigenphase within the modified effective range theory (MERT) \cite{Spurch_1960} framework. MERT and Morrison approaches are, however, applicable only to non-polar interactions. 

In this letter, we propose the estimation of the positron binding energy with polar molecules by employing the SCE method using model correlation potential to perform scattering calculations. The $S$-matrix from the scattering calculations is then continued into the complex momentum plane by fitting it to the Padé approximant to find its pole position and hence the binding energy. 

\textit{Scattering calculation:} The symmetry-adapted SCE approach is implemented to perform the scattering calculations. Here we will present only the key aspects of the method, the details of which can be found in the Refs. \cite{BACCARELLI20111, Winifred, Snigdha_PRA}. In this approach, the potentials, bound
wavefunctions, and continuum wavefunctions are expanded in terms of symmetry-adapted
angular functions (spherical harmonics) centered at a common origin, typically the center of mass of the target
molecule. Eventually, by integrating over the angular grid, we get a set of coupled radial equations for the scattering process,  

\begin{align}
 \left[\frac{d^2}{dr^2}-\frac{l(l+1)}{r^2}+k^2\right] F_{lh}^{p\mu}(r)= 2\sum_{l'h'}V_{lh,l'h'}^{p\mu}(r)F_{l'h'}^{p\mu}(r),
 \label{eqn4}
\end{align}
\textcolor{black}{with $F_{lh}^{p\mu}$ representing the continuum radial wavefunction for the positron.} \textcolor{black}{$k^2/2$ corresponds to the positron energy, $p$ is one of the irreducible representations (IRs) of the molecular point group, and $\mu$ is one of the components of the IR. The index $h$ denotes a basis function for a given partial wave $l$.} These radial equations are solved to get the scattering parameters, viz., $K$- and $T$-matrix elements in the body-fixed frame. Here, $V_{lh,l'h'}^{p\mu}(r)$ are the coupling potential elements given as

\begin{align}
 V_{lh,l'h'}^{p\mu}(r) = \int X_{hl}^{p\mu}(\theta,\phi)V(\textbf{r})X_{h'l'}^{p\mu}(\theta,\phi)d\theta d\phi  
\end{align}
\begin{align}
\text{and}, \ V(\textbf{r}) = V_{st} + V_{corr,pol} \nonumber.
\end{align}

The static potential ($V_{st}$) is calculated exactly using the molecular electron density $\rho(\textbf{r})$, whereas the model proposed by Swann and Gribakin \cite{Swann_Gribakin_2018} is considered for the short-range correlation ($V_{corr}$) potential. The model takes into account the other correlations, like virtual positronium formation, and is given as         

\begin{align}
    V_{corr} (\textbf{r})= - \sum_{A=1}^{N_a} \frac{\alpha_A}{2|\textbf{r}-\textbf{r}_A|^4}\left[1-\exp\left(-\frac{|\textbf{r}-\textbf{r}_A|^6}{\rho_A^6}\right)\right].
\end{align}

Here, \textcolor{black}{$N_a$ is the number of atoms in the molecule and} $\alpha_A$ is the hybrid polarizability of the individual atoms, $A$, as given by Miller \cite{Miller}. It accounts for the chemical environment of the atoms in the molecule. $\rho_A$ is the cut-off radius whose value is obtained for each atom by fitting the computed cross section to the available experimental or \textit{ab initio} data, as elaborated in Appendix \ref{appA}. $\textbf{r}$ denotes the position vector of the positron and $\textbf{r}_A$ is the position vector of the atom $A$. Please note that this model correlation potential will be referred to as $\rm{V_{sg}}$ in this work. Further, for the long-range polarization potential, the given form, which depends upon the \textcolor{black}{dipolar} polarizability ($\alpha$) of the target molecule \cite{SCElib} is used,

\begin{align}
  V_{pol} (\textbf{r})  = - \frac{\alpha}{2r^4},
\end{align}
where $\alpha$ = $\sum_A \alpha_A$. There is no exchange potential involved in the positron scattering. The molecular geometry is optimized, using the Density Functional Theory (DFT) with Perdew–Burke–Ernzerhof (PBE) exchange-correlation functional along with the aug-cc-pVDZ basis set, in the quantum chemistry software package GAUSSIAN 16 \cite{g16}. 

\textit{S-matrix continuation into complex momentum plane:} The calculated $K$- and $T$-matrix elements in the body-fixed frame are transformed to the space-fixed frame employing the POLYDCS code \cite{Polydcs}, along with the ASYMTOP code \cite{JAIN1983301} to generate the asymmetric molecules' rotational energy levels and eigenfunctions. The $s$-wave $T$-matrix element ($T_{00}$) for the  $\rm{J=0\rightarrow0}$ rotational transition (rotationally elastic) is then transformed to the $s$-wave $S$-matrix element ($S_{00}$) using the relation: $S_{00}=1+T_{00}$. It has been shown that the inclusion of rotation for a molecule having a dipole moment greater than the critical value (1.625 D) reduces the infinite number of possible bound states to a few or even zero \cite{Fabrikant_2016}. In the case of weak rotational coupling---like the $\rm{J=0\rightarrow0}$ transition---the poles in the $S$-matrix typically lie on the imaginary $k$ axis, with the ones on the positive axis representing bound states and the ones on the negative axis representing virtual states \cite{Hill_1996}. The analytic continuation of the $S$-matrix element ($S_{00}$)---calculated for the real momentum values ($k$)---into the complex momentum plane is performed by fitting it to the Padé approximant \cite{Chilcott_2021}, which is a rational fraction of two finite-order ($N, M$) polynomials

\begin{align}
S_{00}(k)=\frac{a_0 + a_1k + a_2k^2 + ... + a_Nk^N}{1 + b_1k + b_2k^2 + ... + b_Mk^M}.
\label{eq5}
\end{align}

The linear least-squares fitting procedure is used to determine the values of the $a$ and $b$ coefficients. The denominator of Eq. \ref{eq5} is then equated to zero to find the roots of the polynomial and hence the poles of the $S$-matrix. From the pole position ($k_B$), the binding energy is calculated using: $\varepsilon_B=\frac{k^2_B}{2}$. Please note that the estimated $k_B$ is not purely imaginary but a complex number, of which the imaginary part is used to calculate $\varepsilon_B$, with the real part tolerance kept below 0.05 a.u.    

\textit{Results}: The momentum values for which $S_{00}$ are calculated ranges from 0.00192 a.u. to 0.257 a.u. For the Padé approximant of order $N,M$, the number of data points must be $\ge N+M+1$ to determine the corresponding coefficients. Then, solving the polynomial in the denominator, equated to zero, will result in $M$ number of roots. Not all the $M$ roots will correspond to a real bound state. In order to distinguish the physical bound states from the unphysical ones, we repeated the calculations for $N=M=2$ to $N=M=6$ and filtered out the roots appearing consistently (within a range) for every order. To further strengthen our calculation, for every polynomial order we scanned all the possible data windows to extract the most consistent roots and hence the binding energies. The binding energies determined across all the orders of the Padé approximant for the molecules are summarized in Table \ref{tab:table1}. Table \ref{tab:table1a} compares our calculated $\varepsilon_B$ (we took the average of the $\varepsilon_B$ values in Table \ref{tab:table1}) with the latest experimental result reported by Danielson \textit{et al.} \cite{Danielson_2025} and the \textit{ab initio} MBT estimation of the Hofierka group \cite{hofierka2022many, Baidoo_2024}. The dipole moment and polarizability of the targets, used in the calculations, are also listed in Table \ref{tab:table1a}. 

\textit{Discussion}: There is an overall decent agreement of our calculated $\varepsilon_B$ values, for all the presented molecules, with the experimental results of Danielson \textit{et al.} \cite{Danielson_2025}. For methanol and acetaldehyde, the reported $\varepsilon_B$ values even lie within the experimental uncertainty. For others, though the $\varepsilon_B$ values lie outside the experimental uncertainty, the results are quite promising. For acetone, acetaldehyde, benzaldehyde, and propanal, the MBT results of the Hofierka group \cite{hofierka2022many, Baidoo_2024} are in good agreement with our estimated $\varepsilon_B$ values, except furan, for which our result is quite high. We hope our estimated binding energy can be further improved by incorporating the anisotropy of the atomic hybrid polarizability tensor in the model correlation potential \cite{Swann_2021}. Hence, a model-potential-based SCE approach---which is computationally less demanding than \textit{ab initio} methods---   employing a correlation potential that accounts for many-body correlation, such as virtual positronium formation, can result in a fairly good estimation of $\varepsilon_B$, comparable to experimental and \textit{ab initio} results. Finally, in future work, we aim to employ this simpler and cost-effective approach to predict, through careful analysis, the positron binding energy for systems where experimental data are unavailable.

\textit{Acknowledgements}: Dhanoj Gupta (DG) acknowledges the Science and Engineering Research Board (SERB), Department of Science and Technology (DST), Government of India (Grant No. SRG/2022/000394) for providing a computing facility. This work was financially supported by Vellore Institute of Technology (VIT), Vellore, under the Faculty Seed Grant (RGEMS) (Sanction Order No.: SG20250016).

\textit{Data availability}: The data are available from the authors upon reasonable request.

\begin{table*}
\centering
\begin{ruledtabular}
\begin{tabular}{l c c c c c }
\multirow{2}{*}{Target} & \multicolumn{5}{c}{$\varepsilon_B$ in meV} \\
\cline{2-6} 
 & $N=M=2$ & $N=M=3$ & $N=M=4$ & $N=M=5$ & $N=M=6$  \\
\colrule
Methanol($\rm{CH_3OH}$) & 6.396 & 5.607 & 5.296 & 4.854 & 4.803  \\
Ethanol ($\rm{C_2H_5OH}$) & 25.582 & 24.873 & 27.930 & 25.934 & 23.796  \\
Acetone ($\rm{CH_3COCH_3}$) & 151.838 & 162.522 & 156.143 & 160.417 & 167.762  \\
Acetophenone ($\rm{C_6H_5COCH_3}$) & 225.671 & 265.035 & 220.123 & 290.654 & 318.278  \\
Acetaldehyde ($\rm{CH_3CHO}$) & 84.669 & 92.361 & 89.108 & 89.801 & 90.645                \\
Benzaldehyde ($\rm{C_6H_5CHO}$) & 197.312 & 195.536 & 242.339 & 207.511 & 211.783                \\
Propanal ($\rm{C_2H_5CHO}$) & 90.325 & 108.142 & 103.649 & 105.110 & 101.160                \\
Butanal ($\rm{C_3H_7CHO}$) & 130.025 & 127.171 & 132.170 & 138.877 & 144.911               \\
Furan ($\rm{C_4H_4O}$) & 73.099 & 72.580 & 73.545 & 74.448 & 79.653               \\
\end{tabular}
\label{table1}
\caption{\label{tab:table1} Estimated binding energies ($\varepsilon_B$) for different orders of the Padé approximant.}
\end{ruledtabular}
\end{table*}

\begin{table*}
\centering
\begin{ruledtabular}
\begin{tabular}{l c c c c c c}

Target & $N_e$ & $\mu$ (D)  & $\alpha$ (a.u.) & $\varepsilon_B$ (meV) & $\varepsilon_B^{expt}$ (meV) & $\varepsilon_B^{MBT}$ (meV)   \\ \hline

Methanol ($\rm{CH_3OH}$)  & 18  & 1.588$^*$ & 21.911 & 5.3912 & $6 \pm 1$ & --  \\ 

Ethanol ($\rm{C_2H_5OH}$)  & 26 & 1.513$^*$ & 34.295 & 25.623 & $30 \pm 1$ & -- \\ 

Acetone ($\rm{CH_3COCH_3}$) & 32 & 2.88$^\dagger$ & 42.96 & 159.7364 & $170 \pm 8$ & 152 \\ 
Acetophenone ($\rm{C_6H_5COCH_3}$) & 64 & 3.133$^*$ & 95.768 & 263.9522 & $288 \pm 5$ & -- \\
Acetaldehyde ($\rm{CH_3CHO}$) & 24 & 2.75$^\dagger$ & 30.576 & 89.3168 & $90 \pm 10$ & 89 \\
Benzaldehyde ($\rm{C_6H_5CHO}$) & 56 & 3.140$^{\dagger\dagger}$ & 83.384 & 210.8962 & $220 \pm 5$ & 213 \\
Propanal ($\rm{C_2H_5CHO}$) & 32 & 2.855$^*$ & 42.96 & 101.6772 & $115 \pm 8$ & 108 \\ 
Butanal ($\rm{C_3H_7CHO}$) & 40 & 2.973$^*$ & 55.344 & 134.6308 & $150 \pm 10$ & -- \\ 
Furan ($\rm{C_4H_4O}$) & 36 & 0.660$^{\dagger\dagger}$ & 50.788 & 74.665 & $52 \pm 5$ & 42\\ 
\end{tabular}
\label{table2}
\caption{\label{tab:table1a} Number of electrons ($N_e$), dipole moment ($\mu$: $^*$quantum chemistry calculation using GAUSSIAN 16 \cite{g16}, $^\dagger$Danielson \textit{et al.} \cite{Danielson_2025}, $^{\dagger\dagger}$experimental \cite{NIST}), polarizability ($\alpha$) \cite{Miller}, present binding energy ($\varepsilon_B$), experimental binding energy ($\varepsilon_B^{expt}$) \cite{Danielson_2025}, and \textit{ab initio} MBT binding energy ($\varepsilon_B^{MBT}$) \cite{hofierka2022many, Baidoo_2024}.}
\end{ruledtabular}
\end{table*}

\clearpage
\bibliography{paper_5}

@PREAMBLE{
 "\providecommand{\noopsort}[1]{}" 
 # "\providecommand{\singleletter}[1]{#1}%" 
}

@article{Paul_Pierre,
  title = {Rapid Annihilations of Positrons in Polyatomic Gases},
  author = {Paul, D. A. L. and Saint-Pierre, L.},
  journal = {Phys. Rev. Lett.},
  volume = {11},
  issue = {11},
  pages = {493--496},
  numpages = {0},
  year = {1963},
  month = {Dec},
  publisher = {American Physical Society},
  doi = {10.1103/PhysRevLett.11.493},
  url = {https://link.aps.org/doi/10.1103/PhysRevLett.11.493}
}

@article{Gilbert_2002,
  title = {Vibrational-Resonance Enhancement of Positron Annihilation in Molecules},
  author = {Gilbert, S. J. and Barnes, L. D. and Sullivan, J. P. and Surko, C. M.},
  journal = {Phys. Rev. Lett.},
  volume = {88},
  issue = {4},
  pages = {043201},
  numpages = {4},
  year = {2002},
  month = {Jan},
  publisher = {American Physical Society},
  doi = {10.1103/PhysRevLett.88.043201},
  url = {https://link.aps.org/doi/10.1103/PhysRevLett.88.043201}
}

@article{Danielson_2025,
  title = {Improved positron-molecule binding energies and estimations using molecular parameters},
  author = {Danielson, J. R. and Arthur-Baidoo, E. and Surko, C. M.},
  journal = {Phys. Rev. A},
  volume = {111},
  issue = {4},
  pages = {042809},
  numpages = {16},
  year = {2025},
  month = {Apr},
  publisher = {American Physical Society},
  doi = {10.1103/PhysRevA.111.042809},
  url = {https://link.aps.org/doi/10.1103/PhysRevA.111.042809}
}

@article{rong_radiochemistry_2023,
  title={Radiochemistry for positron emission tomography},
  author={Rong, Jian and Haider, Achi and Jeppesen, Troels E and Josephson, Lee and Liang, Steven H},
  journal={Nature communications},
  url = {https://www.nature.com/articles/s41467-023-36377-4},
  volume={14},
  number={1},
  pages={3257},
  year={2023},
  publisher={Nature Publishing Group UK London}
}

@article{Gribakin_2010,
  title = {Positron-molecule interactions: Resonant attachment, annihilation, and bound states},
  author = {Gribakin, G. F. and Young, J. A. and Surko, C. M.},
  journal = {Rev. Mod. Phys.},
  volume = {82},
  issue = {3},
  pages = {2557--2607},
  numpages = {0},
  year = {2010},
  month = {Sep},
  publisher = {American Physical Society},
  doi = {10.1103/RevModPhys.82.2557},
  url = {https://link.aps.org/doi/10.1103/RevModPhys.82.2557}
}

@article{hofierka2022many,
  title={Many-body theory of positron binding to polyatomic molecules},
  author={Hofierka, Jaroslav and Cunningham, Brian and Rawlins, Charlie M and Patterson, Charles H and Green, Dermot G},
  journal={Nature},
  volume={606},
  number={7915},
  pages={688--693},
  year={2022},
  url =
  {https://doi.org/10.1038/s41586-022-04703-3},
  publisher={Nature Publishing Group UK London}
}

@article{Rawlins_2023,
  title = {Many-Body Theory Calculations of Positron Scattering and Annihilation in ${\mathrm{H}}_{2}$, ${\mathrm{N}}_{2}$, and ${\mathrm{CH}}_{4}$},
  author = {Rawlins, C. M. and Hofierka, J. and Cunningham, B. and Patterson, C. H. and Green, D. G.},
  journal = {Phys. Rev. Lett.},
  volume = {130},
  issue = {26},
  pages = {263001},
  numpages = {9},
  year = {2023},
  month = {Jun},
  publisher = {American Physical Society},
  doi = {10.1103/PhysRevLett.130.263001},
  url = {https://link.aps.org/doi/10.1103/PhysRevLett.130.263001}
}

@article{Cassidy_2024,
  title = {Many-body theory calculations of positron binding to halogenated hydrocarbons},
  author = {Cassidy, J. P. and Hofierka, J. and Cunningham, B. and Rawlins, C. M. and Patterson, C. H. and Green, D. G.},
  journal = {Phys. Rev. A},
  volume = {109},
  issue = {4},
  pages = {L040801},
  numpages = {6},
  year = {2024},
  month = {Apr},
  publisher = {American Physical Society},
  doi = {10.1103/PhysRevA.109.L040801},
  url = {https://link.aps.org/doi/10.1103/PhysRevA.109.L040801}
}

@article{Baidoo_2024,
  title = {Positron annihilation and binding in aromatic and other ring molecules},
  author = {Arthur-Baidoo, E. and Danielson, J. R. and Surko, C. M. and Cassidy, J. P. and Gregg, S. K. and Hofierka, J. and Cunningham, B. and Patterson, C. H. and Green, D. G.},
  journal = {Phys. Rev. A},
  volume = {109},
  issue = {6},
  pages = {062801},
  numpages = {14},
  year = {2024},
  month = {Jun},
  publisher = {American Physical Society},
  doi = {10.1103/PhysRevA.109.062801},
  url = {https://link.aps.org/doi/10.1103/PhysRevA.109.062801}
}

@article{fedus2025semi,
  title={Semi-empirical prediction of bound and virtual states in low-energy positron and electron scattering by atoms and molecules},
  author={Fedus, Kamil and Karwasz, Grzegorz},
  journal={The European Physical Journal D},
  volume={79},
  number={5},
  pages={43},
  year={2025},
  url = 
  {https://doi.org/10.1140/epjd/s10053-025-00994-z},
  publisher={Springer}
}

@article{Swann_Gribakin_2018,
    author = {Swann, A. R. and Gribakin, G. F.},
    title = {Calculations of positron binding and annihilation in polyatomic molecules},
    journal = {The Journal of Chemical Physics},
    volume = {149},
    number = {24},
    pages = {244305},
    year = {2018},
    month = {12},
    issn = {0021-9606},
    doi = {10.1063/1.5055724},
    url = {https://doi.org/10.1063/1.5055724},
}

@article{Suzuki_2020,
  title = {Positron binding in chloroethenes: Modeling positron-electron correlation-polarization potentials for molecular calculations},
  author = {Suzuki, Haruya and Otomo, Takuma and Iida, Ryusei and Sugiura, Yutaro and Takayanagi, Toshiyuki and Tachikawa, Masanori},
  journal = {Phys. Rev. A},
  volume = {102},
  issue = {5},
  pages = {052830},
  numpages = {9},
  year = {2020},
  month = {Nov},
  publisher = {American Physical Society},
  doi = {10.1103/PhysRevA.102.052830},
  url = {https://link.aps.org/doi/10.1103/PhysRevA.102.052830}
}

@article{Swann_Gribakin_2019,
  title = {Positron Binding and Annihilation in Alkane Molecules},
  author = {Swann, A. R. and Gribakin, G. F.},
  journal = {Phys. Rev. Lett.},
  volume = {123},
  issue = {11},
  pages = {113402},
  numpages = {6},
  year = {2019},
  month = {Sep},
  publisher = {American Physical Society},
  doi = {10.1103/PhysRevLett.123.113402},
  url = {https://link.aps.org/doi/10.1103/PhysRevLett.123.113402}
}

@article{Swann_Gribakin_2020,
  title = {Model-potential calculations of positron binding, scattering, and annihilation for atoms and small molecules using a Gaussian basis},
  author = {Swann, A. R. and Gribakin, G. F.},
  journal = {Phys. Rev. A},
  volume = {101},
  issue = {2},
  pages = {022702},
  numpages = {22},
  year = {2020},
  month = {Feb},
  publisher = {American Physical Society},
  doi = {10.1103/PhysRevA.101.022702},
  url = {https://link.aps.org/doi/10.1103/PhysRevA.101.022702}
}

@article{Swann_2021,
  title = {Effect of chlorination on positron binding to hydrocarbons: Experiment and theory},
  author = {Swann, A. R. and Gribakin, G. F. and Danielson, J. R. and Ghosh, S. and Natisin, M. R. and Surko, C. M.},
  journal = {Phys. Rev. A},
  volume = {104},
  issue = {1},
  pages = {012813},
  numpages = {15},
  year = {2021},
  month = {Jul},
  publisher = {American Physical Society},
  doi = {10.1103/PhysRevA.104.012813},
  url = {https://link.aps.org/doi/10.1103/PhysRevA.104.012813}
}

@article{Spurch_1960,
  title = {Modification of Effective-Range Theory in the Presence of a Long-Range Potential},
  author = {Spruch, Larry and O'Malley, Thomas F. and Rosenberg, Leonard},
  journal = {Phys. Rev. Lett.},
  volume = {5},
  issue = {8},
  pages = {375--377},
  numpages = {0},
  year = {1960},
  month = {Oct},
  publisher = {American Physical Society},
  doi = {10.1103/PhysRevLett.5.375},
  url = {https://link.aps.org/doi/10.1103/PhysRevLett.5.375}
}

@article{Snigdha_PRA,
  title = {Effect of correlation on the elastic scattering of slow positrons from molecules},
  author = {Sharma, Snigdha and Gupta, Dhanoj},
  journal = {Phys. Rev. A},
  volume = {113},
  issue = {2},
  pages = {022814},
  numpages = {18},
  year = {2026},
  month = {Feb},
  publisher = {American Physical Society},
  doi = {10.1103/8qlm-nfn6},
  url = {https://link.aps.org/doi/10.1103/8qlm-nfn6}
}

@article{Frighetto_2026,
  title = {Positron scattering by oxygen-containing molecules: Ab initio and model-potential approaches via the Schwinger multichannel method},
  author = {Frighetto, Francisco F. and Goche, Maria Gabrielle and Barbosa, Alessandra Souza and Sanchez, Sergio d'A.},
  journal = {Phys. Rev. A},
  volume = {113},
  issue = {5},
  pages = {052806},
  numpages = {13},
  year = {2026},
  month = {May},
  publisher = {American Physical Society},
  doi = {10.1103/scjg-sjr5},
  url = {https://link.aps.org/doi/10.1103/scjg-sjr5}
}

@article{Frighetto_2024_Low,
  title = {Low-energy positron scattering by saturated and unsaturated hydrocarbons: Cross sections and bound states},
  author = {Frighetto, Francisco Fernandes and Sanchez, Sergio d'Almeida and Barbosa, Alessandra Souza},
  journal = {Phys. Rev. A},
  volume = {110},
  issue = {2},
  pages = {022805},
  numpages = {13},
  year = {2024},
  month = {Aug},
  publisher = {American Physical Society},
  doi = {10.1103/PhysRevA.110.022805},
  url = {https://link.aps.org/doi/10.1103/PhysRevA.110.022805}
}

@article{Morrison,
  title = {Interpretation of the near-threshold behavior of cross sections for $e\ensuremath{-}{\mathrm{CO}}_{2}$ scattering},
  author = {Morrison, Michael A.},
  journal = {Phys. Rev. A},
  volume = {25},
  issue = {3},
  pages = {1445--1449},
  numpages = {0},
  year = {1982},
  month = {Mar},
  publisher = {American Physical Society},
  doi = {10.1103/PhysRevA.25.1445},
  url = {https://link.aps.org/doi/10.1103/PhysRevA.25.1445}
}

@ARTICLE{BACCARELLI20111,
   author       = {Isabella Baccarelli and Ilko Bald and Franco A. Gianturco and Eugen Illenberger and Janina Kopyra},
   title        = {Electron-induced damage of DNA and its components: Experiments and theoretical models},
   year         = {2011},
   journal      = {Physics Reports},
   volume       = {508},
   pages        = {1-44},
   doi     ={https://doi.org/10.1016/j.physrep.2011.06.004},
}

@BOOK{Winifred,
   author       = {Winifred M. Huo and Franco A. Gianturco},
   year         = 1995,
   title        = {Computational Methods for Electron—Molecule Collisions},
   publisher    = {Springer New York, NY}
}

@article{Miller,
author = {Miller, Kenneth J.},
title = {Additivity methods in molecular polarizability},
journal = {Journal of the American Chemical Society},
volume = {112},
number = {23},
pages = {8533-8542},
year = {1990},
doi = {10.1021/ja00179a044},

URL = { 
    
        https://doi.org/10.1021/ja00179a044
    
    

},

}

@ARTICLE{SCElib,
   author       = {N. Sanna and G. Morelli and S. Orlandini and M. Tacconi and I. Baccarelli}, 
   title        = {SCELib4.0: The new program version for computing molecular properties in the Single Center Approach},
   year         = {2020}, 
   journal      = {Computer Physics Communications}, 
   volume       = {248}, 
   pages        = {106970},
   doi = {https://doi.org/10.1016/j.cpc.2019.106970},
}

@article{JAIN1983301,
title = {A program to generate the symmetry-adapted rotational eigenfunctions and energy levels for asymmetric top molecules},
journal = {Computer Physics Communications},
volume = {30},
number = {3},
pages = {301-309},
year = {1983},
issn = {0010-4655},
doi = {https://doi.org/10.1016/0010-4655(83)90097-8},
url = {https://www.sciencedirect.com/science/article/pii/0010465583900978},
author = {Ashok Jain and D.G. Thompson}
}

@ARTICLE{Polydcs,
   author       = {N. Sanna and F.A. Gianturco}, 
   title        = {Differential cross sections for electron/positron scattering from polyatomic molecules}, 
   journal      = {Computer Physics Communications}, 
   volume       = {114}, 
   pages        = {142-167}, 
   year         = {1998}, 
   doi = {https://doi.org/10.1016/S0010-4655(98)00091-5},
}

@article{Fabrikant_2016,
doi = {10.1088/0953-4075/49/22/222005},
url = {https://doi.org/10.1088/0953-4075/49/22/222005},
year = {2016},
month = {nov},
publisher = {IOP Publishing},
volume = {49},
number = {22},
pages = {222005},
author = {Fabrikant, Ilya I},
title = {Long-range effects in electron scattering by polar molecules},
journal = {Journal of Physics B: Atomic, Molecular and Optical Physics}
}

@article{Hill_1996,
  title = {Electron-hydrogen fluoride scattering at ultralow electron energies: Possible role of dipole-supported states},
  author = {Hill, S. B. and Frey, M. T. and Dunning, F. B. and Fabrikant, I. I.},
  journal = {Phys. Rev. A},
  volume = {53},
  issue = {5},
  pages = {3348--3357},
  numpages = {0},
  year = {1996},
  month = {May},
  publisher = {American Physical Society},
  doi = {10.1103/PhysRevA.53.3348},
  url = {https://link.aps.org/doi/10.1103/PhysRevA.53.3348}
}

@article{Chilcott_2021,
  title = {Experimental observation of the avoided crossing of two $S$-matrix resonance poles in an ultracold atom collider},
  author = {Chilcott, Matthew and Thomas, Ryan and Kj\ae{}rgaard, Niels},
  journal = {Phys. Rev. Res.},
  volume = {3},
  issue = {3},
  pages = {033209},
  numpages = {9},
  year = {2021},
  month = {Sep},
  publisher = {American Physical Society},
  doi = {10.1103/PhysRevResearch.3.033209},
  url = {https://link.aps.org/doi/10.1103/PhysRevResearch.3.033209}
}

@misc{g16,
author={M. J. Frisch and G. W. Trucks and H. B. Schlegel and G. E. Scuseria and M. A. Robb and J. R. Cheeseman and G. Scalmani and V. Barone and G. A. Petersson and H. Nakatsuji and X. Li and M. Caricato and A. V. Marenich and J. Bloino and B. G. Janesko and R. Gomperts and B. Mennucci and H. P. Hratchian and J. V. Ortiz and A. F. Izmaylov and J. L. Sonnenberg and D. Williams-Young and F. Ding and F. Lipparini and F. Egidi and J. Goings and B. Peng and A. Petrone and T. Henderson and D. Ranasinghe and V. G. Zakrzewski and J. Gao and N. Rega and G. Zheng and W. Liang and M. Hada and M. Ehara and K. Toyota and R. Fukuda and J. Hasegawa and M. Ishida and T. Nakajima and Y. Honda and O. Kitao and H. Nakai and T. Vreven and K. Throssell and Montgomery, {Jr.}, J. A. and J. E. Peralta and F. Ogliaro and M. J. Bearpark and J. J. Heyd and E. N. Brothers and K. N. Kudin and V. N. Staroverov and T. A. Keith and R. Kobayashi and J. Normand and K. Raghavachari and A. P. Rendell and J. C. Burant and S. S. Iyengar and J. Tomasi and M. Cossi and J. M. Millam and M. Klene and C. Adamo and R. Cammi and J. W. Ochterski and R. L. Martin and K. Morokuma and O. Farkas and J. B. Foresman and D. J. Fox},
title={Gaussian 16 {R}evision {C}.01},
year={2016},
note={Gaussian Inc. Wallingford CT}

}

@misc{NIST,
  title     = "Computational chemistry comparison and benchmark database,
               {NIST} standard reference database 101",
  author    = "Johnson, R D",
  publisher = "National Institute of Standards and Technology",
  year      =  2002,
  url = {http://cccbdb.nist.gov/}
}

@article{frighetto2023low,
  title={Low-energy positron scattering by small hydrocarbons: methane, acetylene, ethylene, and ethane molecules},
  author={Frighetto, Francisco F and Sanchez, Sergio d’A and Barbosa, Alessandra Souza},
  journal={The European Physical Journal D},
  volume={77},
  number={12},
  pages={206},
  year={2023},
  doi={https://doi.org/10.1140/epjd/s10053-023-00787-2

},
  publisher={Springer}
}

\appendix

\section{}
\label{appA}
\begin{figure}[htbp]
 \centering
\includegraphics[width=12cm, height=12cm, keepaspectratio]{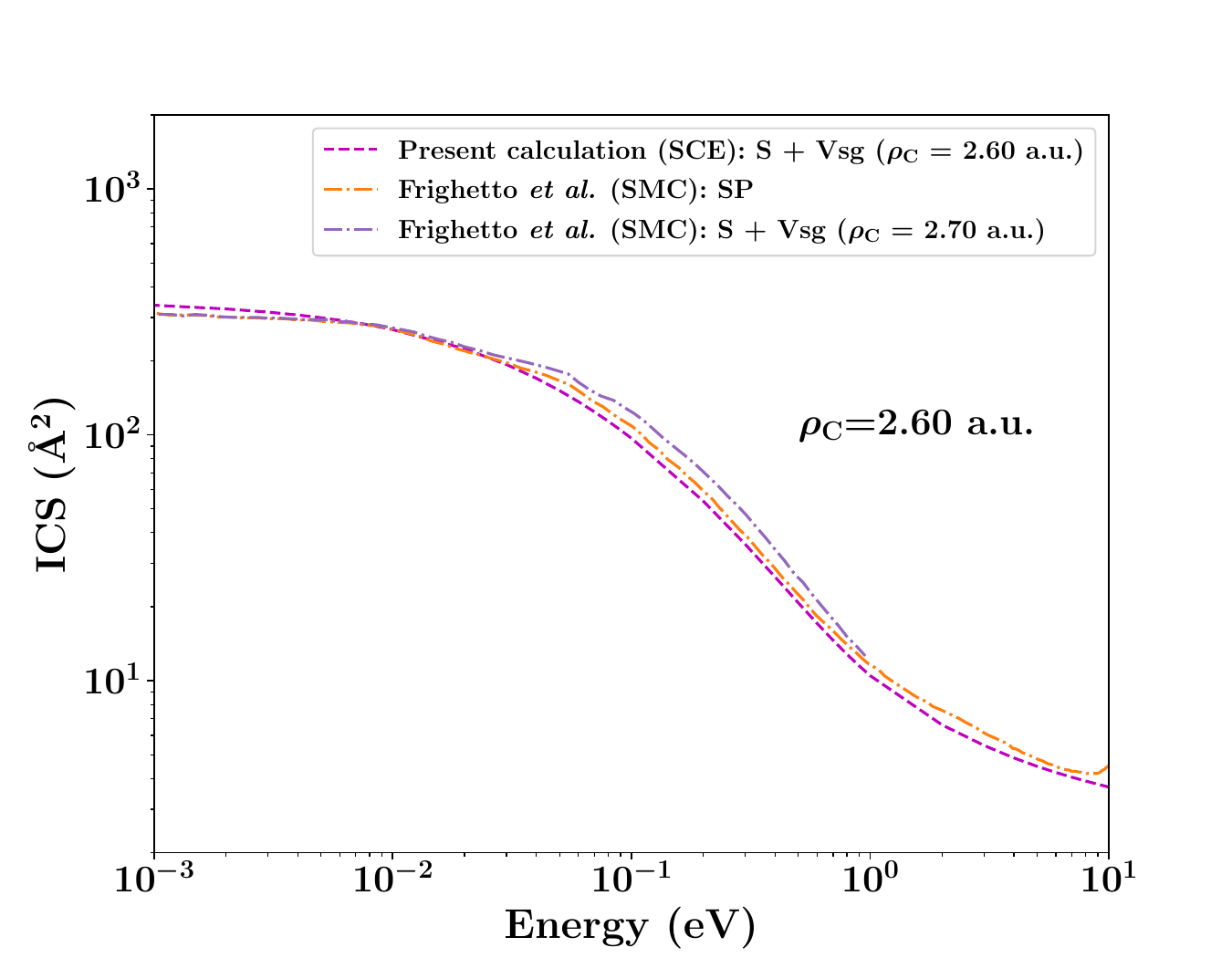}
\caption{\label{fig1}Integral cross section of methane to fix the cut-off radius ($\rho_A$) value for $\rm{sp^3}$ hybridized carbon atom. Magenta dashed line: present calculation using $\rm{V_{sg}}$ correlation. Orange dash-dot line: Frighetto \textit{et al.} \cite{Frighetto_2024_Low,frighetto2023low} SMC-SP data. Purple dash-dot line: Frighetto \textit{et al.} \cite{Frighetto_2024_Low} SMC data using $\rm{V_{sg}}$ correlation.}
\end{figure}

The optimized value of the cut-off radius, $\rho_A$, for different atoms was obtained by directly fitting the calculated integral cross section (ICS) for smaller molecules (containing the desired atoms) with the best available experimental or \textit{ab initio} data. The $\rho_A$ value depended upon the hybridization of the atoms, as listed in Table \ref{table3}. Except for the $\rm{sp^3}$ hybridized carbon atom, the $\rho_A$ values for all other atoms were determined in our previous work \cite{Snigdha_PRA}. $\rho_A$ for the $\rm{sp^3}$ hybridized carbon atom was estimated by fitting the ICS calculated for methane molecule with the \textit{ab initio} SMC calculation of Frighetto \textit{et al.} with the static plus polarization (SP) approximation \cite{Frighetto_2024_Low, frighetto2023low} and the static plus model potential ($\rm{S+V_{sg}}$) approach \cite{Frighetto_2024_Low}, as shown in Fig. \ref{fig1}.

\begin{table}[h]
\centering

\begin{ruledtabular}
\begin{tabular}{c c}

Atom & $\rho_A$ in a.u.  \\ \hline

H & 1.82  \\

C $\rm{(sp)}$ & 1.97 \\ 
C $\rm{(sp^2)}$ & 2.15 \\ 
C $\rm{(sp^3)}$ & 2.60 \\ 
O $\rm{(sp^2)}$ & 1.00 \\ 
O $\rm{(sp^3)}$ & 1.30 \\ 

\end{tabular}
\caption{\label{table3}Cut-off radius ($\rho_A$) determined for the constituent atoms of the targets.}
\end{ruledtabular}
\end{table}

\end{document}